\documentclass[a4paper, twocolumn, 11pt]{quantumarticle}
\pdfoutput=1
\usepackage[utf8]{inputenc}
\usepackage[english]{babel}
\usepackage[T1]{fontenc}
\usepackage{graphicx}
\usepackage{enumitem}
\usepackage{array}
\usepackage{amsmath} 
\usepackage{amsthm}
\usepackage{amssymb}    
\usepackage{latexsym}
\usepackage{amsfonts}
\usepackage{mathrsfs}
\usepackage{color}
\usepackage{bbold}
\usepackage{kotex}
\usepackage[numbers, sort&compress]{natbib}

\newcommand{\bra}[1]{\left< #1\right|}   
\newcommand{\ket}[1]{\left|#1\right>}

\begin{document}

\title{On the feasibility of quantum teleportation protocols implemented with Silicon devices}

\author{Junghee Ryu}
\affiliation{Center for Quantum Information R\&D, Korea Institute of Science and Technology Information (KISTI), \\Daejeon 34141, Republic of Korea}
\affiliation{Division of Quantum Information, KISTI School, Korea University of Science and Technology, \\Daejeon 34141, Republic of Korea}

\author{Hoon Ryu}
\affiliation{Department of Artificial Intelligence Engineering, Kumoh National Institute of Technology, \\Gumi, Gyeongsangbuk-do 39177, Republic of Korea}
\email{electronic mail: elec1020@kumoh.ac.kr}

\maketitle

\begin{abstract}
With recent experimental advancements demonstrating high-fidelity universal logic gates and basic programmability, Silicon-based spin quantum bit (qubit) have emerged as promising candidates for scalable quantum computing. However, implementation of more complex quantum information protocols with many qubits still remains a critical challenge for realization of practical programmability in Silicon devices. In this study, we present a computational investigation of entanglement-based quantum information applications implemented on an electrically defined quantum dot structure in Silicon. Using in-house multi-scale simulations based on tight-binding calculations augmented with bulk physics, we model a five quantum dot system that can create up to five electron spin qubits, and discuss details of control engineering needed to implement single-qubit rotations and two-qubit logic operations in a programmable manner. Using these elementary operations, then, we design a five-qubit quantum teleportation protocol and computationally verify its end-to-end operation including a simple but clear analysis on how the designed circuit can be affected by charge noise. With engineering details that are not well uncovered by experiments, our results demonstrate the advanced programmability of Silicon quantum dot systems, delivering the practical guidelines for potential designs of quantum information processes based on electrically defined Silicon quantum dot structures.

\end{abstract}
%\flushbottom

\section{Introduction}
With its exceptionally long coherence times and compatibility with cutting-edge industrial fabrication technologies~\cite{kobayashi2021engineering, muhonen2014storing, veldhorst2014addressable, kawakami2016gate, bertrand2016fast}, a gate-driven Silicon (Si) quantum dot (QD) system has been regarded as a highly promising platform for scalable quantum logic devices. Another potential strength that the Si QD system has compared to other physical platforms such as superconductors and trapped ions, is the feasibility of integrations of classical control hardwares in a Si wafer, so the platform can also become a strong candidate for chip-level implementation of large-scale processing units. Given these advantages, there have been substantial experimental and modeling efforts aimed at developing practical logic gate operations for electron-spin quantum bits (qubits) in Si QD platforms, where the confinement is controlled by magnetic fields and man-made electrodes~\cite{zajac2018resonantly, takeda2021quantum, philips2022universal, kang2024quantum, ryu2022devitalizing, kang2021exploring, kawakami2014electrical, takeda2016fault, yoneda2018quantum, huang2019fidelity, xue2022quantum, mkadzik2022precision, noiri2022fast, watson2018programmable, takeda2022quantum}. In particular, elaborated device designs for realizing universal logic gates including single qubit rotations and the Controlled-NOT (CNOT) gate, which are fundamental building blocks for various quantum circuits, have been intensively investigated. Consequently, the single qubit rotations can be implemented with a high fidelity larger than $99\%$~\cite{veldhorst2014addressable, kawakami2016gate, kawakami2014electrical, takeda2016fault, yoneda2018quantum, huang2019fidelity, xue2022quantum, takeda2021quantum, mkadzik2022precision, noiri2022fast}. The fidelity of experimentally reported CNOT gates still remains a bit lower than that of single-qubit gates \cite{zajac2018resonantly, takeda2021quantum, watson2018programmable}, but the successful realization of a fast CNOT gate, which is implemented with a single microwave pulse and takes less than 200 nanoseconds (ns) to complete its logic operation, has been demonstrated using a double QD (DQD) platform~\cite{zajac2018resonantly}.

Experimentalists also have put huge efforts on the implementation of circuit-level operations using electron spins in the Si QD platform. Starting from the work reported by Watson $et$ $al$.~\cite{watson2018programmable}, where programmable two-qubit operations are physically verified against the well-known Deutsch-Josza and Grover search algorithm, remarkable advancements have been achieved in terms of the qubit size of programmable circuit units. A three-qubit Greenberger-Horne-Zeilinger (GHZ) entangled state is generated by deploying sequential two-qubit entangling operations in the Si triple QD system~\cite{takeda2021quantum}, and the latest work reported by Philips $et$ $al$., successfully extended the scope of physical implementation of a GHZ resource to a six QD system~\cite{philips2022universal}. While it is indeed true that these achievements indicate significant progresses in designs of scalable quantum processors with a Si QD platform, physical designs and control engineering of many QD systems in Si still remain as
critical challenges for realization of programmable quantum circuits in an algorithmic level, motivating studies based on systematic computer-aided simulations that can be useful to provide practical guideline for device designs, uncovering crucial control parameters for device engineering that often require too huge manpower and time to be secured directly from experiments.

In this work, we computationally investigate designs of a five-qubit quantum circuit using an electrically defined QD system in Si. Extending our previous works that focused on designs of elementary gate operations in a Si DQD system \cite{kang2021exploring, ryu2022devitalizing, kang2024quantum}, here we model the five QD (FQD) structure with our in-house multi-scale modeling approach based on a tight-binding theory augmented with bulk physics, to present details of device designs and control engineering needed to implement a quantum teleportation circuit based on an entanglement swapping protocol that has been proposed with photonic and superconductor qubits \cite{pirandola2015nature} but not yet with semiconductor platforms. Using a realistically sized Si FQD structure as a simulation target, we first find a set of standard operation modes in terms of electrical biases that need to implement single- \& two-qubit gate operations in a programmable manner, and present modeling results on a fully operating quantum teleportation circuit together with an entanglement swapping protocol that involve up to five qubits. Since charge noise is omnipresent in solid systems \cite{kranz2020, wilen2021}, its effect on the fidelity of output states is also examined to discuss reliability of the implemented teleportation circuit. With engineering details that are in principle hard to be predicted with $ab$-$initio$ simulations, this work serves as a rare modeling study on a quantum information protocol that has not been physically realized with Si devices, presenting a practical guideline for potential designs of circuit-based quantum algorithms with a Si QD platform. 

\section{Methods}
\begin{figure*}[t]
\centering
\includegraphics[width=\textwidth]{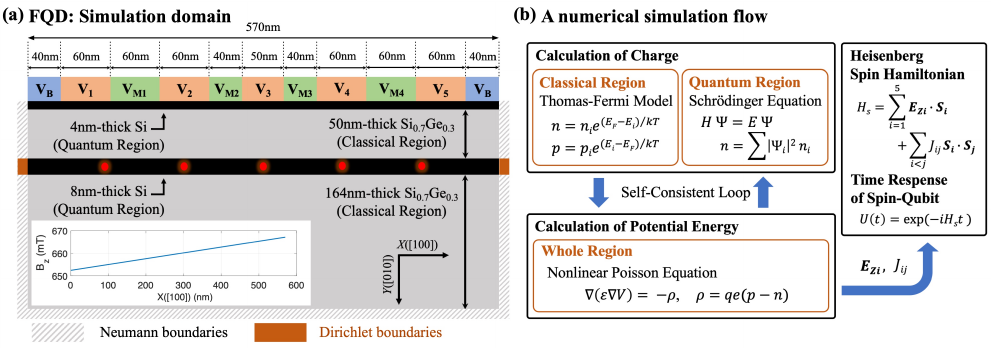}
\caption{
\textbf{Domain description \& Modeling approach:} (a) A two-dimensional simulation domain is employed to describe the Silicon (Si) five quantum dot (FQD) structure that is assumed to be infinitely long along the Z ($[001]$)-direction. Appropriate controls of the electrical biases imposed on top electrodes can create an array of QDs in the middle $8$nm-thick Si layer that confines up to five electron spins. A static magnetic field $B_{Z}$, whose spatial distribution is given in the inset, is applied to make the Zeeman-splitting energy of each electron spin be distinguishable. (b) The modeling process consists of the two steps: (i) charge and potential profiles are self-consistently determined with device simulations based on bulk physics and tight-binding calculations. (ii) The results of device simulations are used to construct a Heisenberg spin Hamiltonian that is solved to investigate the time-dependent behaviors of electron spins in the FQD system.
}
\label{fig:domain}
\end{figure*}

A simulation domain for the Si FQD system is shown in Figure~\ref{fig:domain}(a), which is much larger along the X ([100])-direction than the domain we previously employed to describe the DQD system~\cite{ryu2022devitalizing}. The target system is based on a Si$/$Si-Germanium (SiGe) heterostructure so electrons can be naturally confined in the 8nm-thick middle Si layer along the Y ([010])-direction due to the conduction band offset between Si and Si layers. The system has a total of eleven top electrodes (five leads ($V_{1}, \dots, V_{5}$) with six barrier leads ($V_{M1}, \dots, V_{M4}$, 2$\times$$V_{B}$)) that can be controlled to create up to five potential valleys along the lateral (X) direction in the middle Si layer. A static magnetic field ($B_{Z}$) is applied along the Z ([001])-direction with a gradient of $\sim$0.028 mT$/$nm along the X-direction as shown in the inset of Figure~\ref{fig:domain}(a), which is done with a micromagnet in reality \cite{watson2018programmable, zajac2018resonantly} and makes the electron ground state in each QD have a distinguishable Zeeman-splitting energy. Source \& drain leads (the two X-boundary regions described with a Dirichlet boundary condition) are assumed to be grounded, and $V_{B}$ is set to 200 mV for all the simulations conducted in this work. The FQD structure is described in a two-dimensional manner with a periodic boundary condition along the Z-direction, because physical QD structures that have been reported so far are quite long ($>$ 100 nm) along that direction \cite{zajac2018resonantly, watson2018programmable, takeda2021quantum, philips2022universal} and thus 3D simulations do not necessarily have to be conducted.

Figure~\ref{fig:domain}(b) shows the entire flow of numerical simulations employed in this work, which consists of the following two steps: (i) First, we conduct device simulations for self-consistent evaluation of bias-dependent electrostatic charge \& potential profiles in the FQD system. Here, once the charge profile is given, the corresponding potential distribution is calculated with a nonlinear Poisson equation. The charge density profile, which is determined by the potential distribution, is evaluated in a multi-scale manner to reduce computational cost. That is, the charge profile in the vicinity ($\pm$4 nm) of the middle Si layer (denoted as Quantum Region) is obtained with electronic structure calculations based on a parabolic effective mass model \cite{wang2005}, and the charge profile in the other region (denoted as Classical Region) is obtained analytically with bulk physics since there will be almost no free electrons so quantum-mechanical effects will not play a critical role. Once the step (i) is completed, (ii) we construct a Heisenberg spin Hamiltonian \cite{russ2018prb} with results of device simulations, and solve the corresponding time-dependent Schr\"{o}dinger equation to predict the time response of electron spin qubits. 

\section{Results and Discussion}
\begin{figure*}[t]
\centering
\includegraphics[width=\textwidth]{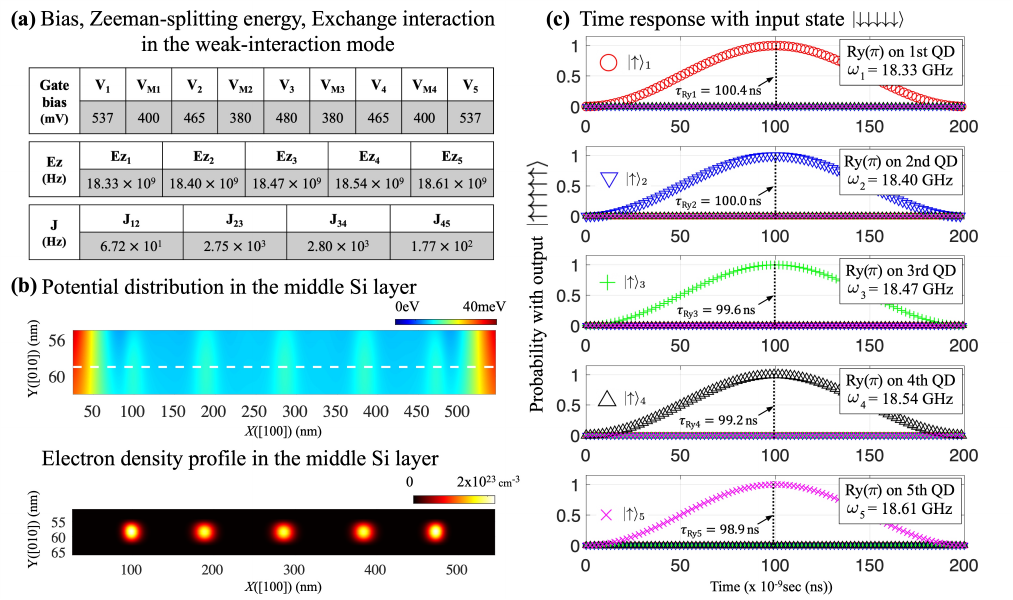}
\caption{
\textbf{System initialization \& Individual qubit addressing:} (a) The set of gate biases that initializes the five quantum dot (FQD) system in a weak-interaction mode is shown with corresponding Zeeman-splitting energies ($E_Z$) and exchange interactions ($J$) between neighboring QDs. The leftmost and rightmost barrier gates are set to $200$ mV ($V_{B}$ in Figure \ref{fig:domain}(a)). At the given set of biases, $J$'s become several kHz or lower, implying the independent controllability of each spin in the FQD system. (b) The upper and lower subfigure show the potential and charge profile in the middle Si layer, respectively. The potential profile indicate the set of gate biases (in Figure \ref{fig:initialization}(a)) satisfies the symmetric biasing condition. (c) Time responses of spins in our FQD system are calculated with a time-varying magnetic field $B_Y(t)$ $=$ $B_{O} \cos \omega_{D} t$ ($B_{O}$ = 5.0 MHz, $\omega_{D}$ = the Zeeman-splitting energy of the electron ground state in each QD). Results clearly show the Y-rotation ($R_Y$) of each spin can be controlled individually with $\omega_{D}$.
}
\label{fig:initialization}
\end{figure*}

The first step for implementation of logic operations with our FQD system is to find a DC bias condition that initializes spin qubit states, where the qubit $\ket{0}$ and $\ket{1}$ are encoded to the electron down-spin ($\ket{\downarrow}$) and up-spin ($\ket{\uparrow}$) ground state of each QD, respectively. As described in the Methods section, applying appropriate DC biases can create the five potential valleys in the middle Si layer and a set of bias conditions must be elaborately designed to ensure that a single electron occupies the down-spin ($\ket{\downarrow}$) ground state in each potential valley (QD). For this purpose, we conducted device simulations extensively and found that all the five $\ket{\downarrow}$ states are occupied with a single electron at ($V_{1}, V_{M1},V_{2}, V_{M2},V_{3}, V_{M3},V_{4}, V_{M4},V_{5}$) = (537 mV, 400 mV, 465 mV, 380 mV, 480 mV, 380 mV, 465 mV, 400 mV, 537 mV), ensuring initialization of the FQD system to a five-qubit state $\ket{\downarrow \downarrow \downarrow \downarrow \downarrow}$. At this bias condition, exchange interactions between neighboring $\ket{\downarrow}$ states turn out to be quite weak so, as summarized in Figure~\ref{fig:initialization}(a), the interaction energy $J_{12}$, $J_{23}$, $J_{34}$, and $J_{45}$ ($J_{ij}$ = the interaction between $i$-th and $j$-th QDs) marks 6.72$\times 10^{1}$ Hz, 2.75$\times 10^{3}$ Hz, 2.80$\times 10^{3}$ Hz, and 1.77$\times 10^{2}$ Hz, respectively, where $J$ of a DQD system was reported to be $\sim$ 7.54$\times 10^4$ Hz in the weak-interaction mode~\cite{ryu2022devitalizing}. The five Zeeman-splitting energies $E_{Z1}$, $E_{Z2}$, $E_{Z3}$, $E_{Z4}$, and $E_{Z5}$ ($E_{Zi}$ = the Zeeman-splitting energy of the $i$-th QD) are calculated as 18.33 GHz, 18.40 GHz, 18.47 GHz, 18.54 GHz, and 18.61 GHz, respectively. Figure~\ref{fig:initialization}(b) shows the corresponding potential distribution and charge density profile in the middle Si layer, indicating the secured biases also satisfies a symmetric bias condition ($i.e.$ the potential distribution is symmetric along the X-direction) that is reported to be generally helpful to increase noise-robustness of the QD system \cite{zajac2018resonantly, reed2016}.

\begin{figure*}[t!]
\centering
\includegraphics[width=\textwidth]{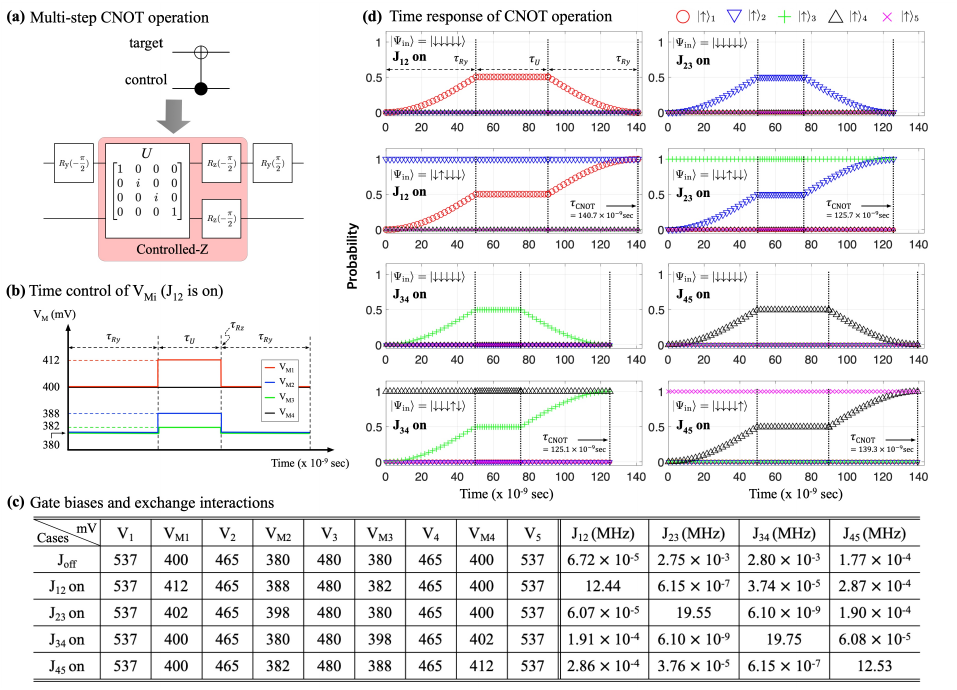}
\caption{
\textbf{Programmable implementation of multi-step CNOT gating:} (a) In this work, we implement the Controlled-NOT (CNOT) logic with three steps: (i) a Y-rotation of the target spin by -$\pi/2$ radian ($R_{Y}(-\pi/2)$), (ii) a Controlled-Z (CZ) operation with neighboring spins, and (iii) $R_{Y}(+\pi/2)$. The CZ logic can be implemented only with DC biases that must be different from the one secured for a weak-interaction mode. (b) The exchange interaction between neighboring spins can be controlled by adjusting the barrier gates ($V_{Mi}$), and the graph shows the time-sequence of $V_{M1}$-$V_{M4}$ that must be applied to implement a CNOT logic with spins in the leftmost two quantum dots (QD1 and QD2). (c) The table summarizes gate biases that make all the five QDs interact weakly ($J_{\text{off}}$), or the spins in $i$-th and $j$-th QD interact strongly ($J_{ij}$ on). (d) The time responses of spins calculated with different sets of $V_{Mi}$'s clearly show that the CNOT logic is conducted in a programmable manner.
}
\label{fig:cnot}
\end{figure*}

With the set of DC biases given in Figure~\ref{fig:initialization}(a), individual qubit addressing can be conducted since the FQD system is in a weak interaction mode. To model the single-qubit rotation logic, we compute the time response of spin states by solving a time-dependent Schr\"{o}dinger equation described with the Heisenberg spin Hamiltonian, where the [010]-oriented time-varying magnetic pulse $B_{Y} (t) = B_{O} \cos(\omega_{D} t+\theta)$, a control signal needed to make each spin rotate, is considered as elements of the Hamiltonian. In Figure~\ref{fig:initialization}(c), we show the time response of a $\ket{\uparrow}$ state in each QD that is calculated against the input state $\ket{\downarrow \downarrow \downarrow \downarrow \downarrow}$ with $B_{O}$ = 5.0 MHz, $\theta$ = 0 radian, and the driving frequency $\omega_{D}$ = the Zeeman-splitting energy of each QD ($E_{Zi}$). Since the inhomogeneous static magnetic field $B_{Z}$ makes the Zeeman-splitting energies $E_{Zi}$ distinguishable, the individual spin rotation can be controlled by setting $\omega_{D}$ equal to $E_{Zi}$, so, for instance, when $\omega_{D}$ = $E_{Z1}$ (18.33 GHz), the spin rotation around the Y-axis happens only in the $1$st QD and the spin qubit in the rest QDs remains unchanged. According the our modeling results in Figure \ref{fig:initialization}(c), it takes 100.11 ns to complete rotation of the spin in QD1 by $\pi$ radian ($R_Y(\pi)$ gating), and 100.02 ns, 99.92 ns, 99.80 ns, and 99.72 ns with the spin in QD2, QD3, QD4, and QD5, respectively. The progressively faster completion times for QD2, QD3, QD4, and QD5 compared to QD1 are due to the inhomogeneity of $B_{Z}$ that directly affects the driving frequency of a time-varying control pulse $B_{Y} (t)$. The fidelity of a single-qubit $R_Y(\pi)$ operation marks 99.99\% for all the five QDs.

Implementation of two-qubit entangling logics requires strong exchange interaction between two spins where the resonance frequency of one spin state becomes affected by the status of the other spin state. The CNOT gate, which is the most important entangling logic for circuit-based quantum computing, has been a target of physical realization using a Si QD platform, and Zajac $et$ $al.$ implemented a CNOT operation with a single microwave pulse, demonstrating the completion of gating within 200 ns~\cite{zajac2018resonantly}. However, recently it has been shown that this single-step CNOT logic is vulnerable to charge noise in Si devices, and multi-step implementation with a Controlled-Z (CZ) logic is much more robust to charge noise~\cite{ryu2022devitalizing}. Taking the multi-step CNOT logic as a target of programmable designs with a linear array of FQDs, we use the scheme of implementation that requires a sequential conduction of a $R_{Y}(-\pi/2)$ gate $\rightarrow$ a CZ gate $\rightarrow$ a $R_{Y}(\pi/2)$ gate, as shown in Figure~\ref{fig:cnot}(a). Here, the $R_{Y}(\pm \pi/2)$ operations applied to the target qubit should be performed in a weak-interaction mode, and the CZ block, which generates two-qubit entanglement, requires a strong-interaction mode. Once we get the appropriate biasing condition, the CZ gate can be easily implemented with a sequential conduction of a two-qubit Ising ZZ gate with the phase $\pi/2$ and two $R_{Z} (-\pi/2)$ rotations that are native the Si QD platform ($i.e.$ can be implemented with no time-varying pulses)~\cite{ryu2022devitalizing, russ2018prb}. Figure~\ref{fig:cnot}(b) shows a time-varying profile of the four $V_{Mi}$ biases that drive a multi-step CNOT operation using QD1 and QD2 in our FQD system. In the first stage where $R_{Y}(\pi/2)$ is conducted to the spin in QD1, $V_{Mi}$'s are set to the values of the weak-interaction mode (Figure~\ref{fig:initialization}(a)). Once the $R_{Y}(\pi/2)$ gating is completed, $(V_{M1}, V_{M2}, V_{M3}, V_{M4})$ need to be switched to (412 mV, 388 mV, 382 mV, 400 mV) where $J_{12}$ reaches to the order of MHz ($\sim$1.24$\times$10${^7}$ Hz, being reasonably strong compared to the experimentally reported results~\cite{takeda2021quantum, philips2022universal}) but $J_{23}$,  $J_{34}$ and  $J_{45}$ still remain weak. Upon completion of the CZ operation, $(V_{M1}, V_{M2}, V_{M3}, V_{M4})$ need to be back to their weak-interaction values in order that another single qubit rotation $R_{Y}(-\pi/2)$ can be conducted.

Figure~\ref{fig:cnot}(c) summarizes the computationally secured device controls that initialize the FQD system to a $\ket{\downarrow \downarrow \downarrow \downarrow \downarrow}$ state in the following modes that are necessary to implement the CZ operation in a programmable manner (note that $V_1$-$V_5$ do not change): (i) All the five exchange interactions are weak ($J_{\text{off}}$), (ii) only $J_{12}$ reaches to the order of MHz ($J_{12}$ on), (iii) only $J_{23}$ reaches to the order of MHz ($J_{23}$ on), (iv) only $J_{34}$ reaches to the order of MHz ($J_{34}$ on), and (v) only $J_{45}$ reaches to the order of MHz ($J_{45}$ on). Simulated time responses of spin states in the FQD system, given in Figure~\ref{fig:cnot}(d), clearly support programmable operations of a multi-step CNOT logic so, when $J_{12}$ is on, the spin in QD1 (red, target qubit) is flipped ($\ket{\downarrow}_{1} \rightarrow \ket{\uparrow}_{1}$) only when the spin in QD2 is $\ket{\uparrow}_{2}$ (blue, control qubit), completing the CNOT operation in $\sim$140.7 ns. This conditional spin-flip operation applies to other cases as well, so the CNOT operation for a QD pair of 2-3, 3-4, and 4-5 are completed in 125.7 ns, 125.1 ns, 139.3 ns, respectively. Variation in the operation time here is mainly because $J_{23}$ (19.6 MHz) and $J_{34}$ (19.8 MHz) are bigger than $J_{12}$ (12.4 MHz) and $J_{45}$ (12.5 MHz) when they are on. In all the simulation cases, we assumed that transitions of DC biases are conducted instantaneously, and the fidelity of CNOT gating reaches 99.99\%.

\begin{figure*}[t!]
\centering
\includegraphics[width=\textwidth]{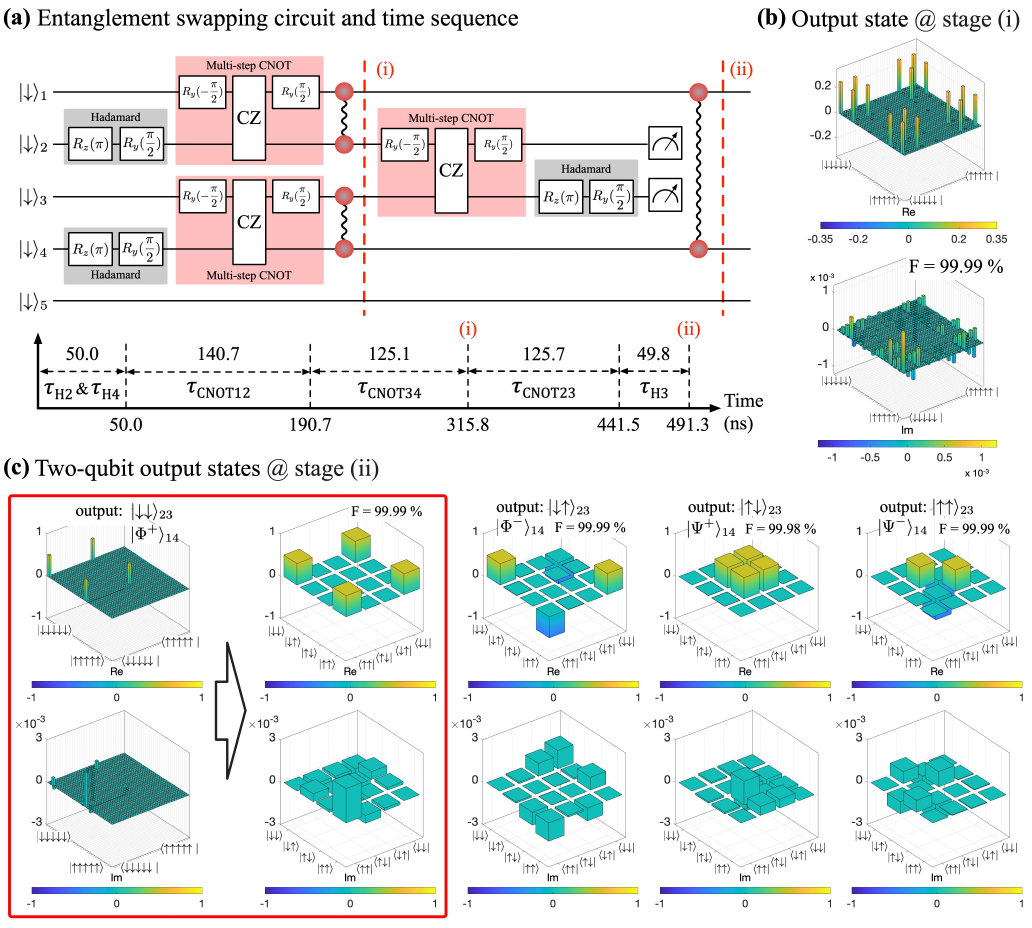}
\caption{
\textbf{Entanglement swapping protocol:} (a) A entanglement swapping circuit designed with our Silicon five quantum dot (QD) system is presented in a logical level with operation times of logic blocks. Generation of the two Bell states is completed in $315.8$ ns (up to stage (i)) and the entire swapping process is completed in $491.3$ ns excluding measurement operations. (b) A density matrix of the output state at stage (i) shows that two Bell states are created for (QD1, QD2) and (QD3, QD4) with a fidelity of $99.99 \%$ in a noise-free condition. (c) Density matrices of output states at stage (ii) are shown, where each case corresponds to the four possible outcomes of measurement conducted against spins in QD2 \& QD3. For clear visualization, full density matrices are redrawn against two relevant qubits (QD1 \& QD4) as shown by the process enclosed with a red solid line.}
\label{fig:swapping}
\end{figure*}

Based on the secured controls for programmable implementation of single-qubit rotations and a two-qubit CNOT logic, now we explore designs of a quantum teleportation module using the FQD system, and, for this purpose, we begin the discussion from implementation of an entanglement swapping circuit that prepares remote quantum entanglement between subsystems that have never directly interacted, serving as a key element for realization of long-distance quantum communications~\cite{zukowski1993event, bose1998multiparticle, dur1999quantum}. Presenting a novel method for transmission of quantum information that cannot be copied and pasted due to the no-cloning theorem \cite{wootters1982}, quantum teleportation uses an entanglement swapping protocol as a channel to surpass the capability of classical communications. So here we first implement the entanglement swapping protocol to generate remote quantum entanglement, and then will employ the output state (swapped entangled state) of the swapping protocol 
to realize quantum teleportation. The process of entanglement swapping using a five-qubit circuit, obviously, should begin with preparation of two independent Bell states: one between QD1 and QD2, and the other between QD3 and QD4. These Bell states are represented as $\ket{\Phi^{+}}_{12} \otimes \ket{\Phi^{+}}_{34}$, where each $\ket{\Phi^{+}}_{ij} = \frac{1}{\sqrt{2}}(\ket{\downarrow \downarrow}_{ij}+\ket{\uparrow \uparrow}_{ij})$ denotes the maximally entangled state made with $i$-th and $j$-th QD in the system. To generate these Bell states, we first apply a Hadamard gate to the spin in QD2 and QD4, creating an equal superposition state of $\ket{\uparrow}$ and $\ket{\downarrow}$ in both QD2 and QD4 separately. Subsequently, a multi-step CNOT operation
needs to be conducted to the spins in QD1-QD2 and QD3-QD4 pairs, utilizing the spins in QD2 and QD4 as control qubits and those in QD1 and QD3 as target qubits. As depicted in Figure~\ref{fig:swapping}(a) that presents a circuit representation of the swapping protocol, Hadamard gates here is implemented with a sequential conduction of $R_Y$ and $R_Z$ gates that takes $\sim$50 ns in the weak-interaction mode ($J_{\text{off}}$, see Figure~\ref{fig:cnot}(c)). As discussed with Figure~\ref{fig:cnot}(a) and Figure \ref{fig:cnot}(b), the CNOT operation involves the strong-interaction mode, and the two CNOTs shown in Figure~\ref{fig:swapping}(a) cannot be conducted simultaneously since they must be conducted with a different set of $V_{Mi}$'s. Upon completion of the two CNOT operations, we reach the stage (i) in Figure~\ref{fig:swapping}(a), getting the tensor product of Bell states $\ket{\Phi}_{12} \otimes \ket{\Phi}_{34}$. Our results indicate the entire process up to stage (i) is completed in 315.8 ns, and the operation times for sub-logic components belonging to this process are also presented in the inset of Figure~\ref{fig:swapping}(a). Figure~\ref{fig:swapping}(b) shows the density matrix calculated for the output state at stage (i). Exhibiting a total of sixteen clear peaks in the real part with a value of 0.25, the density matrix confirms a characteristic signature of the Bell states with a fidelity of $99.99\%$.

\begin{figure*}[t!]
\centering
\includegraphics[width=\textwidth]{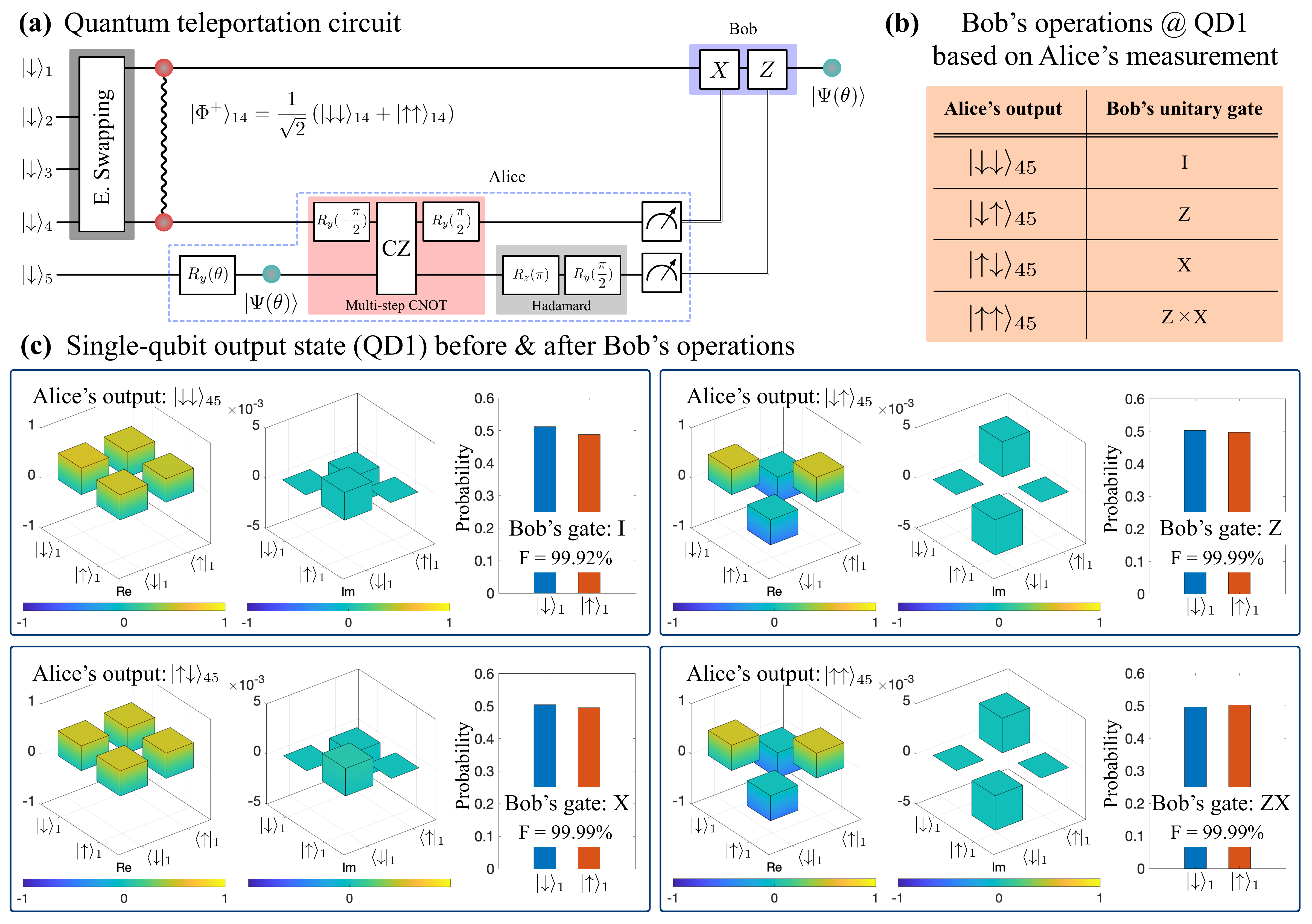}
\caption{
\textbf{Quantum teleportation module:} (a) A quantum teleportation module implemented with our five quantum dot (QD) system is shown in a logical level. Once the entanglement swapping process (Figure \ref{fig:swapping}(a)) is completed, its output state is utilized as a quantum channel to teleport an arbitrary single-qubit state that is prepared by the sender (Alice) with a Y-rotation ($R_Y(\theta)$) of the spin in QD5. This single-qubit state, together with the spin of QD4, is then subjected to a Bell measurement. (b) Depending on the outcome of a Bell measurement, the receiver (Bob) conducts a predefined set of logic operations to the spin of QD1 to get the final outcome. (c) Density matrices of the spin state in QD1 prior to Bob's single-qubit operations are presented. Bar graphs show the spin probabilities of QD1 that are obtained after conduction of Bob's logic operations.
}
\label{fig:teleportation}
\end{figure*}

\begin{figure*}[t!]
\centering
\includegraphics[width=\textwidth]{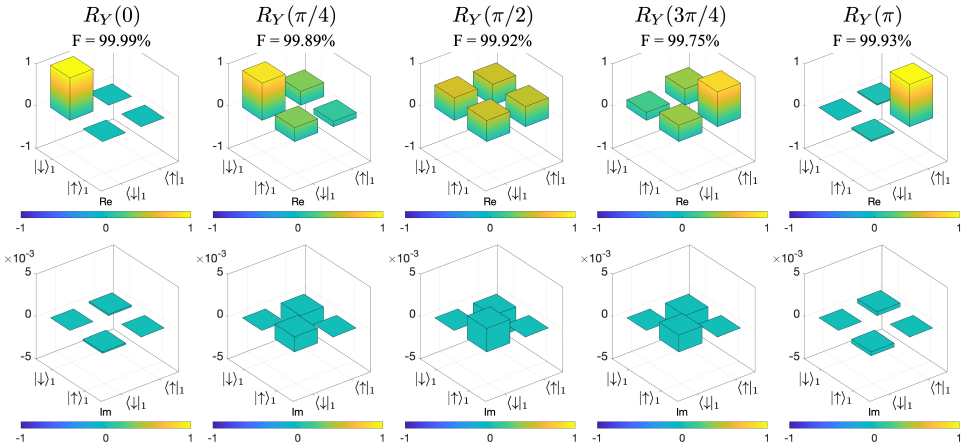}
\caption{
\textbf{Quantum teleportation of arbitrary single-qubit states:} The teleportation module in Figure \ref{fig:teleportation}(a) is simulated for various angles ($\theta$) of the Y-rotation that Alice conducts to the spin in QD5, and density matrices presented here show the final spin qubit in QD1 when the outcome of a Bell measurement is $\ket{\downarrow \downarrow}_{45}$. The high fidelity ($99.75\%$-$99.99\%$) obtained in a noise-free condition supports the precise operation of the implemented module for arbitrary single-qubit input states.
}
\label{fig:teleportation_arbi}
\end{figure*}

Once $\ket{\Phi}_{12} \otimes \ket{\Phi}_{34}$ is successfully secured, the next step of entanglement swapping is to conduct a Bell measurement on the spins in QD2 and QD3. The Bell measurement, which is generally a joint measurement performed on a two-qubit state, can be effectively implemented using computational basis supplemented by appropriate quantum operations that in this case consist of a CNOT gate applied to the spins in QD2 and QD3 followed by conduction of a Hadamard gate on the spin in QD3. The measurement outcomes of the spins in QD2 and QD3 directly determine the two-qubit output states characterized by QD1 and QD4, projecting them into one of the four Bell states. To mimic the measurement process of the computationally obtained output states, here we employ four specific projection operators, each corresponding to one of the possible outcome bases: $\bf 1 \otimes \ket{\downarrow}\bra{\downarrow} \otimes \ket{\downarrow}\bra{\downarrow} \otimes \bf 1 \otimes \bf 1$, $\bf 1 \otimes \ket{\downarrow}\bra{\downarrow} \otimes \ket{\uparrow}\bra{\uparrow} \otimes \bf 1 \otimes \bf 1$, $\bf 1 \otimes \ket{\uparrow}\bra{\uparrow} \otimes \ket{\downarrow}\bra{\downarrow} \otimes \bf 1 \otimes \bf 1$, and $\bf 1 \otimes \ket{\uparrow}\bra{\uparrow} \otimes \ket{\uparrow}\bra{\uparrow} \otimes \bf 1 \otimes \bf 1$ ($\bf 1$ = a single-qubit identity operator). The output entangled state obtained at stage (ii) in Figure~\ref{fig:swapping}(a) is presented in Figure~\ref{fig:swapping}(c), where we represented the relevant two-qubit entangled states for clarity (QD2 and QD3), reducing the originally calculated five-qubit states as shown in the left density matrix enclosed with a red box. According to the four possible measurement outcomes of the spins in QD2 and QD3 $(\ket{\downarrow \downarrow}_{23}$, $\ket{\downarrow \uparrow}_{23}$, $\ket{\uparrow \downarrow}_{23}$, and $\ket{\uparrow \uparrow}_{23})$, results clearly exhibit four distinct patterns of peaks in the real part of the density matrix. Indicating the completion of a swapping process, each pattern of peaks here confirms the successful creation of one of the four Bell states as follows:
\begin{eqnarray*}
\ket{\downarrow \downarrow}_{23} &\rightarrow& \ket{\Phi^{+}}_{14} = \frac{1}{\sqrt{2}} \left( \ket{\downarrow \downarrow}_{14} + \ket{\uparrow \uparrow}_{14} \right), \nonumber \\
\ket{\downarrow \uparrow}_{23} &\rightarrow& \ket{\Phi^{-}}_{14} = \frac{1}{\sqrt{2}} \left( \ket{\downarrow \downarrow}_{14} - \ket{\uparrow \uparrow}_{14} \right), \nonumber \\
\ket{\uparrow \downarrow}_{23} &\rightarrow& \ket{\Psi^{+}}_{14} = \frac{1}{\sqrt{2}} \left( \ket{\downarrow \uparrow}_{14} + \ket{\uparrow \downarrow}_{14} \right), \nonumber \\
\ket{\uparrow \uparrow}_{23} &\rightarrow&  \ket{\Psi^{-}}_{14} = \frac{1}{\sqrt{2}} \left( \ket{\downarrow \uparrow}_{14} - \ket{\uparrow \downarrow}_{14} \right).
\label{eq:fourBell}
\end{eqnarray*}
%\begin{align*}
%\ket{\downarrow \downarrow}_{23} &\rightarrow \ket{\Phi^{+}}_{14} = \frac{1}{\sqrt{2}} \left( \ket{\downarrow \downarrow}_{14} + \ket{\uparrow \uparrow}_{14} \right),
%& \ket{\downarrow \uparrow}_{23} \rightarrow \ket{\Phi^{-}}_{14} = \frac{1}{\sqrt{2}} \left( \ket{\downarrow \downarrow}_{14} - \ket{\uparrow \uparrow}_{14} \right), \\
%\ket{\uparrow \downarrow}_{23} &\rightarrow \ket{\Psi^{+}}_{14} = \frac{1}{\sqrt{2}} \left( \ket{\downarrow \uparrow}_{14} + \ket{\uparrow \downarrow}_{14} \right),
%& \ket{\uparrow \uparrow}_{23} \rightarrow  \ket{\Psi^{-}}_{14} = \frac{1}{\sqrt{2}} \left( \ket{\downarrow \uparrow}_{14} - \ket{\uparrow \downarrow}_{14} \right).
%\label{eq:fourBell}
%\end{align*}
The swapping protocol designed with the Si FQD system here works quite well, so the fidelity of finally generated Bell states reaches $\sim$ 99.9\% consistently for all the four cases. According to our modeling results, it takes $\sim$ 491.3 ns until the designed protocol completes entanglement swapping process excluding measurement, as Figure~\ref{fig:swapping}(a) shows.

Quantum teleportation requires a shared entangled state between the sender (Alice) and the receiver (Bob) that will serve as a quantum channel \cite{ning2019deterministic,pirandola2015nature}. The Bell entangled states generated between the spatially separated spins in QD1 and QD4 from the swapping protocol, which we have discussed so far with Figure \ref{fig:swapping}, can be used to facilitate the transfer of quantum information from Alice to Bob. In our FQD system, a quantum bit to be teleported is encoded to the spin in QD5, and Figure~\ref{fig:teleportation}(a) shows a five-qubit circuit that conducts the teleportation process. Here, the FQD system is first initialized to a $\ket{\downarrow \downarrow \downarrow \downarrow \downarrow}$ state. The entanglement swapping protocol is then employed to create the Bell entangled state between the spins in QD1 and QD4, where the protocol circuit is specifically set to generate the $\ket{\Phi^{+}}_{14}$ state, leaving the spin in QD5 is still left unaffected. Next, Alice prepares the quantum state to be teleported by applying the $R_{Y}(\theta)$ gate operation to the spin in QD5, and, as a result, a superposed state $\ket{\Psi(\theta)}$ (= $\cos(\theta/2)\ket{\downarrow}_{5}$ + $\sin(\theta/2)\ket{\uparrow}_{5}$) is generated in QD5. For implementation of the teleportation, here we set $\theta$ to $\pi/2$, preparing $\ket{\Psi(\pi/2)}$ = $(\ket{\downarrow}_{5} + \ket{\uparrow}_{5})/\sqrt{2}$. The next step is Alice's Bell measurement that is conducted to two distinct qubits (QD4 and QD5), and, as we did for the entanglement swapping protocol (see Figure~\ref{fig:swapping}), the measurement process is mimicked with a sequential conduction of multi-step CNOT and Hadamard operations followed by projective measurements in the computational basis. According to our modeling results, the additional operations needed to implement the teleportation, being enclosed with a blue dashed line in Figure~\ref{fig:teleportation}(a), take about 238.3 ns to be completed (49.5 ns for $R_{Y}(\pi/2)$, 139.3 ns for CNOT, and 49.5 ns for Hadamard), and, including the entanglement swapping time of 491.3 ns, the total operation time of the designed teleportation circuit is $\sim$ 729.6 ns with no consideration of projective measurements in the computational basis.

\begin{figure*}[t]
\centering
\includegraphics[width=\textwidth]{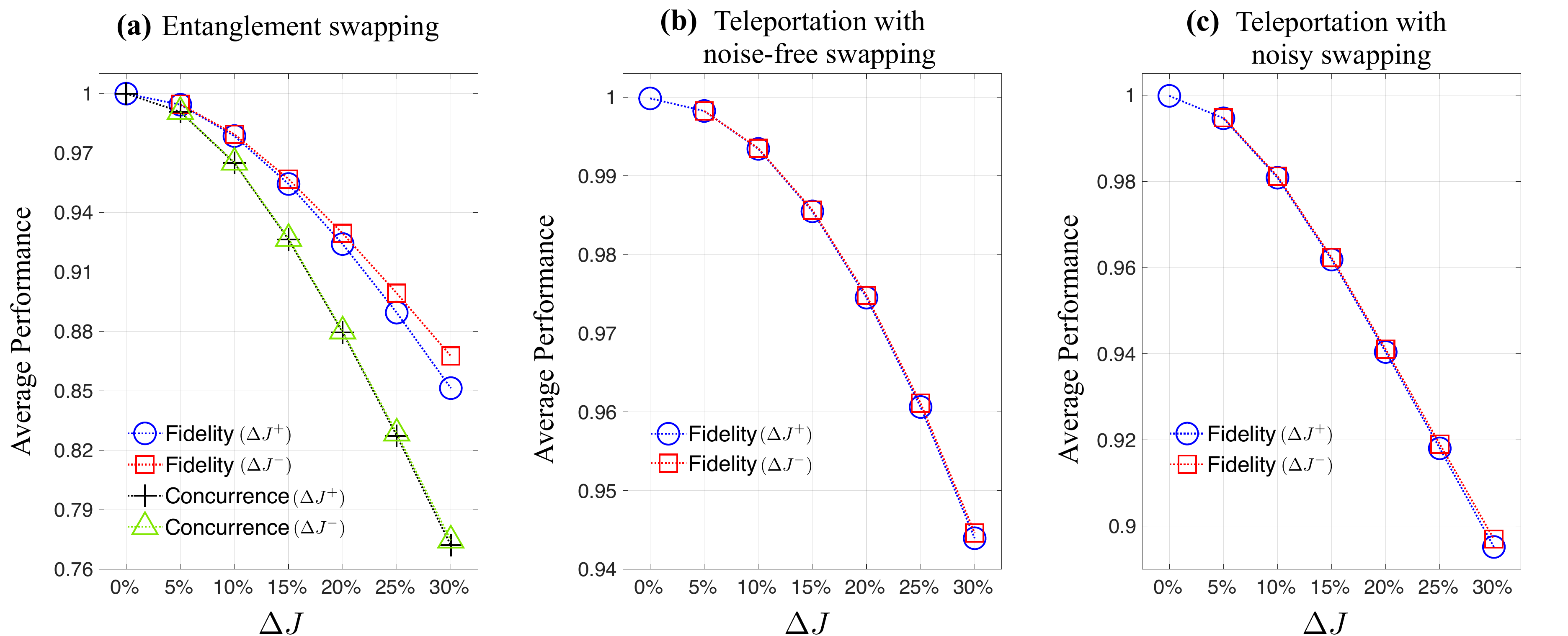}
\caption{
\textbf{Sensitivity of entanglement swapping \& teleportation process to charge noise:} The noise-driven performance degradation is calculated for (a) the entanglement swapping protocol in Figure \ref{fig:swapping}(a), (b) the teleportation module in Figure \ref{fig:teleportation}(a), and (c) the entire circuit operation. The effects of charge noise are incorporated in simulations by introducing variations ($\Delta J$) to exchange interactions secured in a noise-free condition. In all the plots, $\Delta J^{+}$ and $\Delta J^{-}$ indicate positive and negative fluctuations of exchange interactions, respectively. For the entanglement swapping protocol ((a)), the average fidelity and concurrence of output states at stage (ii) (see Figure \ref{fig:swapping}(a)) are plotted as a function of ($\Delta J$). The performance of the teleportation module ((b)) is understood with the average fidelity of the final spin state in QD1 that is calculated assuming a noise-free entanglement swapping process. The noise sensitivity of the entire circuit ((c)) is calculated similarly, but with a noisy swapping protocol. 
}
\label{fig:average}
\end{figure*}

Obviously, the teleportation process must be completed with reception of the information. For this purpose, Bob needs to conduct local operations to the spin qubit in QD1 depending on the outcome of Alice's measurement as summarized in Figure~\ref{fig:teleportation}(b), and, through this final process, the spin qubit state in QD1 can be transformed to the state that Alice intended to teleport. In Figure~\ref{fig:teleportation}(c), we show the density matrices of the spin qubit state in QD1 that are calculated against the four possible outcomes of Alice's measurement before Bob conducts local operations. A bar graph given in the right side of each density matrix shows the probability of ($\ket{\downarrow}$, $\ket{\uparrow}$) in QD1 that represents the output state of Bob's local operation. If teleportation is successful, the output state should precisely match the state Alice intended to send, and modeling results indicate the state received by Bob has a fidelity $>$ 99.9\% against the one sent by Alice in all the cases. Since the results presented in Figure~\ref{fig:teleportation}(c) only explain the case of $\theta$ = $\pi/2$, we conducted more simulations to verify the correct operation of the circuit shown in Figure~\ref{fig:teleportation}(a) for arbitrary $\theta$. Figure~\ref{fig:teleportation_arbi} shows the density matrix of the output state that is calculated with $\theta = (0, \frac{\pi}{4}, \frac{3\pi}{4}, \pi)$ when the outcome of Alice's measurement is $\ket{\downarrow \downarrow}_{45}$. The high fidelity of the output state ranging from 99.75\% to 99.99\% clearly supports accuracy and reliability of the protocol designed with our Si FQD system.

So far, our analysis has been conducted in a noise-free condition. However, quantum circuits in reality are susceptible to performance degradation due to noises. In general, spin qubit operations in Si-based systems suffer from spin or charge noises that reside in solid materials~\cite{kranz2020,maurand2016cmos}. While it is well known that the intrinsic spin noise stemming from non-zero net nuclear spins can be significantly suppressed by using isotopically purified $^{28}$Si where nuclear spin interactions can be hugely eliminated~\cite{muhonen2014storing,maurand2016cmos}. The charge noise, which is due to unexpected fluctuations in electric fields is quite hard to be suppressed in Si devices \cite{kranz2020}, so here we examine the effect of charge noise on the performance of our teleportation process. Since the charge noise mainly affects exchange interactions~\cite{ryu2022devitalizing,russ2018prb,gullans2019prb}, we include it in simulations by introducing a variation to the calculated reference values of exchange interactions as $J \rightarrow J_{\text{ref}} \times (1\pm \Delta J)$, where $\Delta J$ and $J_{\text{ref}}$ represents a small perturbation and a reference value obtained with device simulations (see Figure \ref{fig:cnot}(c)), respectively. First, we calculate the noise-driven performance degradation of the designed entanglement swapping circuit (Figure \ref{fig:swapping}), where the performance is analyzed with the two quantities - the similarity (fidelity) between the entangled states generated under noise and those obtained with $\Delta J$ = 0, and the concurrence of noisy states that explains the entanglement strength~\cite{guhne2009entanglement}. Figure~\ref{fig:average}(a) shows the results, where the average of state fidelities and concurrences calculated for the four distinct Bell states is presented as a function of $\Delta J$ with separate plots for the positive fluctuation ($\Delta J^{+}$) and the negative fluctuation ($\Delta J^{-}$). At $\Delta J = \pm30\%$, which is the worst condition we considered as done by Gullans $et$ $al.$ \cite{gullans2019prb}, the state fidelity marks $0.851$ ($\Delta J^{+}$) and $0.868$ ($\Delta J^{-}$) with the concurrence of $0.772$ ($\Delta J^{+}$) and $0.775$ ($\Delta J^{-}$), respectively. We note that these values are similar to those reported in the case of superconductor qubits, where the state fidelity ranges from $0.872$ to $0.893$ with the concurrence of $0.758$-$0.794$~\cite{ning2019deterministic}. 

We now explore how charge noise affects the entire process of teleportation. For this purpose, we first limit the scope of our noise analysis to the teleportation module shown in Figure~\ref{fig:teleportation}(a), employing one of ideal entangled states as a quantum channel. For each of all the four Bell states, we conduct the teleportation of a single-qubit state prepared at QD5 with $\theta$ = $\pi/2$, and evaluate the fidelity between the teleported state in QD1 and the original state (= $\ket{\Psi(\pi/2)}$). Figure~\ref{fig:average}(b) shows the average fidelity as a function of $\Delta J$, indicating that the fidelity drops to $0.944$ ($\Delta J^{+}$) and $0.945$ ($\Delta J^{-}$) in the worse case. When the noise is incorporated to both the swapping protocol and the teleportation process, the average state fidelity at $\Delta J = \pm 30\%$ becomes $0.895$ and $0.897$ for $\Delta J^{+}$ and $\Delta J^{-}$, respectively. Although simulation results indicate that the average state fidelity and concurrence decreases with increasing strength of noise, the performance of the entire protocol designed with Si QDs still remains above acceptable thresholds for practical quantum teleportation process, because the simulated teleportation fidelity exceeds the classical maximum of $2/3$~\cite{horodecki1999general} and is still higher than the one ($0.55$-$0.95$) reported for other physical platforms \cite{pirandola2015nature}.

\section{Conclusion}
We conducted a comprehensive computational investigation of entanglement-based quantum information applications that are implemented with an electrode-driven Silicon (Si) five quantum dot (FQD) device where quantum bits (qubits) are encoded to spins of confined electrons. Through rigorous modeling of a realistically sized Si/Si-Germanium heterostructure based on bulk physics augmented with tight-binding calculations, we elaborately secured the sets of DC control biases that initialize the Si FQD system, demonstrating implementation of single-qubit rotation and two-qubit Controlled-NOT (CNOT) operation in a programmable manner. To verify the programmability of the Si FQD system in an algorithmic level, we then present an in-depth guideline for design of a quantum teleportation protocol that has not been physically realized yet using Si devices. Starting from implementation of an entanglement swapping circuit that produces a shared entangled state between two spins that do not directly interact, we simulate end-to-end operations of a quantum teleportation protocol that fully utilizes the five spin qubits created in the FQD system, predicting their speed \& accuracy under no noises. Finally, we effectively examine how charge noise affects the teleportation protocol, by incorporating variations to exchange interaction energies that are calculated in a noise-free condition. Results show that the overall performance of our protocol is comparable to or, in some cases, exceeds that reported for other physical platforms, revealing the potential strength of electrically-driven QD systems based on a well-purified Si wafer for implementation of a quantum teleportation circuit.

\section*{Acknowledgements}
This work is supported by the National Research Foundation of Korea grant (NRF-2022M3E4A1072893) funded by the Korea government (MSIT). The NURION supercomputing resource supported by the Korea Institute of Science and Technology Information (KISTI) has been extensively utilized for all the simulations.

\bibliographystyle{unsrt}
\bibliography{FQD_swapping}

\end{document}